\documentclass[cameraready]{Interspeech}
\usepackage{cite}

\title{SAGE: Switch-Aware EEG-Guided Soft Gating for Target Speaker Extraction with In-Trial Switching}

\author[affiliation={1}, orcid=0009-0006-3230-4764]{Xuefei}{Wang}
\author[affiliation={1}, orcid=0009-0004-2463-6094]{Ximin}{Chen}
\author[affiliation={1}, orcid=0009−0009−0716−9988]{Yuting}{Ding}
\author[affiliation={2}]{Chunlin}{Li}
\author[affiliation={1}, orcid=0000-0002-6988-492X, correspondingauthor]{Fei}{Chen}

\address{
    $^1$ Department of Electronic and Electrical Engineering, Southern University of Science and Technology, Shenzhen, China \\
    $^2$ School of Biomedical Engineering, Capital Medical University, Beijing, China
}

\email{12431243@mail.sustech.edu.cn, fchen@sustech.edu.cn}


\usepackage{comment}

\begin{document}

\maketitle
\begin{abstract}
    EEG-guided target speaker extraction is challenging under in-trial auditory attention switching, where neural noise and intrinsic latency can delay or destabilize attention tracking. Conventional methods struggle with dynamic switches and often cause discontinuities at switching points. Therefore, we propose SAGE, a switch-aware EEG-guided soft gating framework that treats in-trial switching as dynamic selection. SAGE generates two candidate speech streams with a robust separator and uses an EEG-guided switch-aware gating module to produce smooth fusion weights and suppress transition artifacts. We further integrate latency-compensated alignment and an uncertainty-driven conservative strategy to handle latency discrepancies and fluctuating EEG reliability. SAGE outperforms baselines, achieving 8.67 dB SI-SDR and 88.24\% STOI while reducing average switching latency to 2.04 s. By coupling neural decoding with speech separation, it enables robust target extraction in dynamic scenarios.

\end{abstract}

\section{Introduction}

The cocktail party problem \cite{Haykin2005the} describes the remarkable human ability to focus on a target talker in complex multi-speaker environments through selective auditory attention. Target speaker extraction (TSE) aims to address this challenge by leveraging auxiliary cues to separate the attended speech signal from a mixture. Electroencephalography (EEG) provides a direct measure of a listener’s neural activity and cognitive state, offering neurophysiological evidence of which speaker is being attended \cite{Mir2016dec, str2021neural}. With millisecond-level temporal resolution, EEG can track the dynamics of attention in real time, making it an attractive physiological signal for guiding target speaker extraction.

Driven by rapid advances in deep learning, the performance of TSE has improved substantially \cite{del2018Asingle, ge2022lspex}. Conventional TSE methods typically rely on enrollment utterances of the target speaker or pre-extracted speaker embeddings, requiring prior access to target speech samples and performing well in fixed single-target settings \cite{xu2020spex,luo2019conv,ji2020spe,ge2020Spex}. To reduce dependence on enrollment speech, researchers have explored alternative auxiliary cues, including spatial information for localization-assisted separation \cite{gu2019neural}, visual cues such as lip movements to enhance directional focus and robustness \cite{ito2021adu}, and biologically inspired mechanisms that emulate neural processing to improve source separation \cite{hos2022end}.

In recent years, EEG-based TSE has emerged as an active research topic. By decoding auditory attention information from EEG, neuro-steered target speaker extraction has shown strong potential for practical applications \cite{wang2025smr}. Pan et al. proposed NeuroHeed+, which significantly improves neuro-guided speaker extraction by jointly modeling auditory attention decoding \cite{Pan2024NeuroHeed}. Fan et al. introduced DGSD, which applies dynamic-graph self-distillation to EEG-based auditory spatial attention decoding and improves the accuracy of feature learning \cite{Fan2024DGSD}. Despite these advances, most existing EEG-guided TSE methods adopt a key simplifying assumption that the listener’s attentional focus remains static during an experimental segment. This assumption overlooks a more realistic setting where listeners may spontaneously switch attention between speakers within a single trial, referred to as in-trial switching.

EEG-guided TSE under in-trial attention switching introduces two central challenges. First, EEG signals are inherently noisy and non-stationary, and are sensitive to environmental interference and inter-subject variability. These properties often degrade the reliability of neural decoding around switching moments, reducing the accuracy of detecting and tracking attention shifts \cite{nog2019to, osu2015att, wang2026eeg}. Second, neural responses exhibit intrinsic latency, leading to inevitable latency discrepancies between attention-switch labels and true switches. When combined with conventional approaches that rely on fixed speaker references and lack mechanisms to accommodate dynamically switching attention, this mismatch often produces delayed responses and audible discontinuities at switching points, which substantially harms continuity and intelligibility \cite{belo2021eeg, geir2021ele, osu2017ne}. Although prior studies have attempted to improve EEG decoding through signal enhancement and feature selection \cite{fug2017noi, som2019neural}, they do not provide dedicated mechanisms tailored to spontaneous in-trial attention switching, and cannot fundamentally address the performance degradation caused by endogenous, dynamic attention switches.

To tackle these issues, we propose SAGE, a switch-aware EEG-guided soft-gating framework for target speaker extraction during spontaneous in-trial speaker switching. SAGE makes three main contributions: (1) it uses a front-end separation module to generate two candidate speech streams for dynamic selection; (2) it designs an EEG-guided switch-aware gating module with adaptive temperature scaling to produce smooth fusion weights and avoid abrupt switching artifacts; and (3) it incorporates latency-compensated alignment with learnable temporal shifts and an uncertainty-driven conservative strategy to handle EEG noise and neural latency. Experiments on spontaneous attention-switching datasets show that SAGE improves speech extraction performance, reduces switching delay, and enhances system stability compared with strong baselines.

\begin{figure*}[th]
  \centering
  \includegraphics[width=17cm]{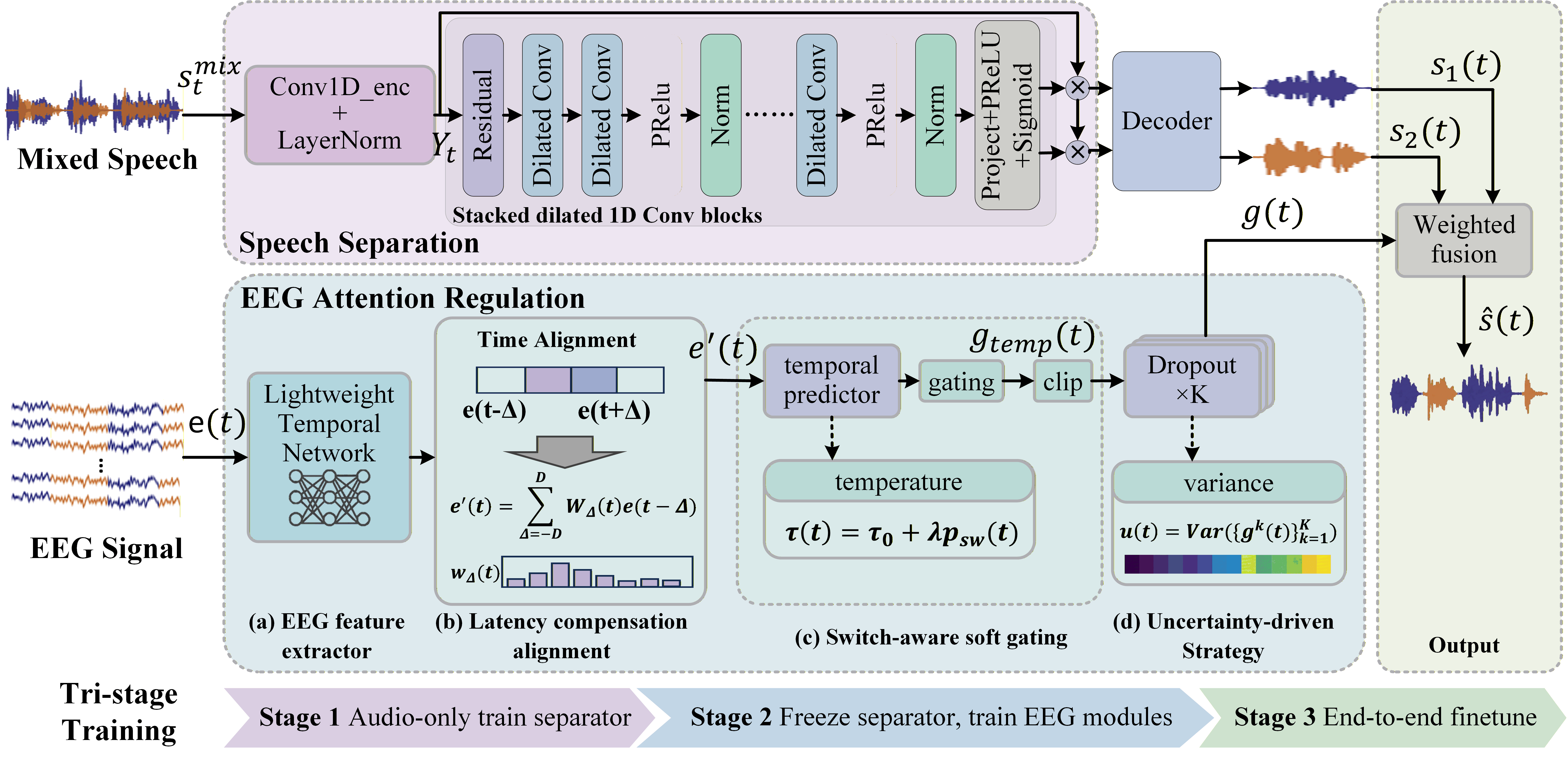}
  \caption{The overall framework of the proposed SAGE, switch-aware EEG-guided soft gating for target speaker extraction.}
  \label{fig:fig1}
\end{figure*}

\section{Method}
\label{sec:methods}
In this section, we propose a novel dynamic target speaker extraction method based on EEG signals, designed to overcome the limitations of traditional speech separation methods in handling in-trial target speaker switching.

\subsection{Architecture}
\label{ssec:2.1}
As shown in Fig. ~\ref{fig:fig1}, the overall framework consists of two main parts: speech separation and EEG attention regulation. The EEG attention regulation module includes EEG feature extractor, switch-aware soft gating, uncertainty-driven conservative strategy, and latency compensation alignment.
The architecture utilizes EEG-guided attention switching information and compensates for neural delays, effectively ensuring smooth transitions during target speaker switching. Given a mixed speech waveform $S^{mix}_t \in \mathbb{R}^{T\times 1}$ and an EEG sequence $e(t)\in \mathbb{R}^{C}$ (with $C$ channels) sampled and synchronized to the audio timeline, the model outputs the extracted target waveform $\hat{s}(t)$ that follows the attended speaker even when attention switches.

The speech feature extraction module follows the Conv-TasNet pipeline~\cite{luo2019conv}, which encodes the time-domain mixture into a learnable latent representation and performs separation by estimating masks in the latent space. Given the input mixture $S_t^{mix}$, a 1-D convolutional encoder first maps it to an embedding sequence, followed by layer normalization:
\begin{equation}
\label{eq:1}
Y_t = \mathrm{LayerNorm}\left(\mathrm{Conv1D}_{\mathrm{enc}}\left(S_t^{mix}\right)\right),
\end{equation}
where $Y_t\in\mathbb{R}^{T'\times D}$ denotes the encoded features, $T'$ is the frame length after convolution, and $D$ is the encoder feature dimension. On top of $Y_t$, a temporal convolutional separator composed of stacked dilated 1-D convolutional blocks aggregates long-range context and predicts two mask sequences through a $1\times1$ projection with PReLU and sigmoid activations. The masks are applied to the mixture embedding to form two candidate latent streams. Finally, a decoder reconstructs the time-domain candidates to produce two candidate speech streams, $s_1(t), s_2(t)$, which provide input for speaker selection and synthesis.

The EEG attention regulation module generates a time-varying weight function $g(t)$ based on the neural activity in the EEG signals, which controls the selection of the final output speech stream. We employ a soft selection mechanism based on a time window, aiming to predict the dynamic change of the target speaker based on the EEG signal. The process is formulated as:
\begin{equation}
\label{eq:2}
\hat{s}(t) = g(t) \cdot s_1(t) + (1 - g(t)) \cdot s_2(t),
\end{equation}
where $g(t) \in [0,1]$ is the dynamic weight generated by the neural network based on the EEG signal, indicating the selected target speaker at time $t$.

\subsection{Switch-aware Soft Gating}
\label{ssec:2.2}
EEG attention switches are often annotated as discrete change points, yet the underlying neural evidence evolves continuously, exhibiting a slowly varying transition rather than an instantaneous jump. Direct hard selection often produces discontinuities or momentary leakage of both speakers around switching points. Therefore, we propose a switch-aware soft gating mechanism that can introduce smoother transitions by incorporating switch probabilities and temperature scaling.

We predict a switch probability $p_{sw}(t)\in[0,1]$ reflecting whether attention is changing around time $t$, and an attention bias logit $\alpha(t)\in\mathbb{R}$ indicating which candidate is more likely attended. Both are produced from the aligned EEG feature via lightweight temporal modeling which enforces temporal consistency. We first compute an intermediate temperature-controlled gate $g_{\mathrm{temp}}(t)$:
\begin{equation}
\label{eq:3}
g_{\mathrm{temp}}(t)=\sigma\left(\frac{\alpha(t)}{\tau(t)}\right), \quad \tau(t)=\tau_0+\lambda\cdot p_{sw}(t),
\end{equation}
where $\tau_0>0$ is a base temperature and $\lambda\ge 0$ controls how much additional smoothing is introduced near potential switches. 
Although $g_{\mathrm{temp}}(t)$ already adapts its sharpness via $\tau(t)$, residual jitter can still arise from EEG noise, especially near switching regions. We therefore obtain the final gating signal $g(t)$ by applying a local diffusion to $g_{\mathrm{temp}}(t)$:
\begin{equation}
\label{eq:4}
g(t)=\text{clip}\Big((g_{\mathrm{temp}} * w_{\sigma(t_0)})(t)\Big),
\end{equation}
where $*$ denotes 1D convolution over time, $w_{\sigma(t_0)}$ is a normalized smoothing kernel centered at a detected switching center $t_0$ with learnable width $\sigma(t_0)>0$, and $\text{clip}(\cdot)$ enforces the range $[0,1]$. The switching centers $\{t_0\}$ are obtained by thresholding peaks or local maxima in $p_{sw}(t)$. 

\subsection{Latency Compensation and Uncertainty Strategy}
\label{ssec:2.3}
In practice, attention-related EEG patterns inevitably lag behind annotated switch events, exhibiting intrinsic latency discrepancies that vary significantly across subjects and trials. Without explicitly compensating for such delays, the gating signal $g(t)$ tends to update after the labeled switch time, increasing wrong-speaker leakage in the transition region. To alleviate this issue, we introduce a differentiable time-alignment module that produces an aligned EEG representation $e'(t)$.

Rather than predicting a single global offset, we formulate alignment as a local, time-varying soft shift within a bounded window:
\begin{equation}
\label{eq:5}
e'(t)=\sum_{\Delta=-D}^{D} w_{\Delta}(t)\cdot e(t-\Delta),
\end{equation}
where $D$ is the maximum shift, $e(t)\in\mathbb{R}^{C}$ denotes the $C$-channel EEG sample at time $t$ and $w_{\Delta}(t)$ is the time-shifting weight dynamically predicted by the network. This operation can be interpreted as differentiable temporal resampling over a plausible neural-latency window, allowing the model to softly aggregate EEG evidence from nearby time points instead of committing to a hard shift. The aligned feature $e'(t)$ is then fed into the gating and switch-prediction network to estimate $\alpha(t)$ and $p_{sw}(t)$, better synchronizing EEG-driven decisions with acoustic changes.
However, EEG quality is not constant, artifacts, subject fatigue, and transient noise can make EEG-derived cues unreliable. Over-confident gating under unreliable EEG can lead to persistent wrong-speaker selection. Therefore, We also incorporate an uncertainty-driven conservative strategy that down-weights aggressive switching and encourages smoother gating when EEG is uncertain.
We estimate uncertainty by performing $K$ stochastic forward passes with dropout enabled in the EEG-related subnetworks, producing a set of gating samples ${g^{(k)}(t)}_{k=1}^{K}$. The uncertainty can be measured as the temporal variance:
\begin{equation}
\label{eq:6}
u(t)=Var\left(\left \{ g^{(k)}(t) \right \}_{k=1}^{K}\right),
\end{equation}
and normalized into $[0,1]$ to obtain a stable control signal. A higher $u(t)$ indicates lower confidence in EEG-driven selection.
When $u(t)$ is high, we increase the preference for continuity by amplifying a smoothness penalty. We implement this by uncertainty-weighted smoothness regularization:
\begin{equation}
\label{eq:7}
L_{\text{smooth}}=\sum_t u(t)\cdot \mathcal{R}\left(g(t)\right),
\end{equation}
where $\mathcal{R}(g(t))$ is a differentiable smoothness regularizer. This design prevents the model from making sharp, low-confidence decisions, while still allowing fast transitions when EEG evidence is both strong and reliable.

\subsection{Loss Function Design}
\label{ssec:2.5}
During training, we optimize the model with a combination of speech reconstruction loss, gating regularization, and switch-structure constraints to improve both extraction quality and temporal stability.

To suppress jitter in $g(t)$, we design a switch-aware variant that relaxes the penalty near predicted switching moments:
\begin{equation}
\label{eq:8}
L_{tv\text{-}sw}=\sum_t \left(1-\beta p_{sw}(t)\right),|g(t)-g(t-1)|,
\end{equation}
where $\beta\in[0,1]$ controls how much smoothing is reduced when $p_{sw}(t)$ is high. This encourages stable gating during steady attention while allowing faster transitions when the model detects a switch. Besides, We supervise the extracted waveform $\hat{s}(t)$ with the ground-truth target speech $s(t)$. Since SI-SDR is a maximization metric, we minimize its negative form.
The final training objective combines these losses, including the uncertainty-weighted smoothness in Eq.~\eqref{eq:7}, with appropriate weights:
\begin{equation}
\label{eq:9}
L = -\mathrm{SI\text{-}SDR}\left(\hat{s}(t), s(t)\right) + \gamma_1 L_{tv\text{-}sw} + \gamma_2 L_{\text{smooth}},
\end{equation}
where $\gamma_1, \gamma_2$ are scalar weights.
To ensure stable convergence, we adopt a tri-stage training strategy. In stage one, we train the audio separator using only audio supervision. In stage two, we partially freeze the separator and jointly train the EEG-related modules. Finally, we perform end-to-end fine-tuning with a smaller learning rate.

\section{Experiments} 
\label{sec:pagestyle}
\subsection{Dataset} 
\label{ssec:Dataset}

In this study, we use a custom-built dataset for spontaneous auditory attention switching based on electroencephalogram (EEG) signals \cite{wang2026neural}. The experimental procedures have been approved by the Ethical Review Board of Southern University of Science and Technology (Approval No. 2022DZX003). The participants consisted of 18 healthy native Mandarin-speaking adults (ages 18-27) with normal hearing. During the experiment, EEG data was recorded using 64 electrodes placed according to the international 10-20 system, with a sampling rate of 500 Hz and mastoid reference electrodes. The auditory stimuli consisted of spatialized mixed speech from one male and one female speakers, presented at $+90^{\circ}$ and $-90^{\circ}$ azimuths. Participants were instructed to switch attention spontaneously and signal the switch using a button press. The experiments were conducted in a soundproof room, and the stimuli were presented through headphones at a sound pressure level of 65 dB SPL. Data preprocessing followed standard procedures, including artifact removal, re-referencing, EEG downsampling to 128 Hz, and application of a 0.1-45 Hz bandpass filter. Additional information and data can be found at: \url{https://doi.org/10.5281/zenodo.17413336}.

\subsection{Experimental Details} 
\label{ssec:Experiment Details}

The proposed model was implemented in Python 3.9 using the PyTorch framework, and all experiments were conducted on a high-performance computing system equipped with NVIDIA V100 GPUs to accelerate training and evaluation. During training, we employed the Adam optimizer with a learning rate of $1 \times 10^{-4}$, a batch size of 16, and default momentum parameters to optimize model stability and convergence. Each participant's dataset was split into training, validation, and test sets in a 8:1:1 ratio to ensure the model was trained, validated, and evaluated on separate data. The training process spanned multiple epochs, and early stopping was employed based on validation loss to prevent overfitting. Additionally, regularization techniques such as dropout and weight decay were applied during training to improve the model's robustness.
During the evaluation phase, we reported scale-invariant signal-to-distortion ratio (SI-SDR), short-time objective intelligibility (STOI), switch detection accuracy (ACC) and average switch latency (ASL).

\section{Results and Discussion}
\subsection{Comparison with Baseline Methods}
To evaluate our EEG-based target speaker extraction approach, we compared it with several baseline models. As shown in Table~\ref{tab:1}, our method, SAGE, outperformed all baselines across all metrics on the spontaneous attention-switching dataset.

SAGE achieved 8.67 dB in SI-SDR, surpassing BASEN (4.02 dB), NeuroHeed (4.96 dB), NeuroSpex+ (6.21 dB), and M3ANet (7.13 dB), demonstrating better interference suppression and speech distortion reduction for higher reconstruction quality. In terms of speech intelligibility, SAGE reached a STOI score of 88.24\%, outperforming BASEN (74.83\%), NeuroHeed (79.57\%), NeuroSpex+ (82.84\%), and M3ANet (84.30\%), preserving perceptually important speech components for clearer, more intelligible extraction.
SAGE achieved the lowest average switching latency of 2.04 s, outperforming BASEN (2.93 s), NeuroHeed (2.81 s), NeuroSpex+ (2.56 s), and M3ANet (2.37 s). This advantage underscored the effectiveness of the latency-compensated alignment module in alleviating the temporal mismatch between EEG attention labels and actual auditory attention switching, enabling faster and more reliable tracking. Overall, the proposed SAGE framework delivered superior performance over multiple strong baselines, with particularly pronounced gains during dynamic target switching.

\begin{table}[!ht]
\renewcommand\arraystretch{1.25}
\caption{Comparison with representative EEG-based target speaker extraction baselines on the spontaneous attention-switching dataset.}
\centering
\scalebox{1.0}{
\begin{tabular}{lcccc}
\hline\hline
\textbf{Methods}         & \textbf{SI\_SDR (dB)}  & \textbf{STOI (\%)}      & \textbf{ASL (s)}         \\ \hline
BASEN~\cite{Zhang2023BASEN}       & 4.02          & 74.83          & 2.93          \\
NeuroHeed~\cite{Pan2024NeuroHeed} & 4.96          & 79.57          & 2.81          \\
NeuroSpex+~\cite{De2025NeuroSpex} & 6.21          & 82.84          & 2.56          \\
M3ANet~\cite{Fan2025M3ANet}       & 7.13          & 84.30          & 2.37          \\
SAGE (proposed)     & \textbf{8.67} & \textbf{88.24} & \textbf{2.04} \\  \hline\hline
\end{tabular}}
\label{tab:1}
\end{table}

\subsection{Ablation Study}

\begin{figure}[th]
  \centering
  \includegraphics[width=8cm]{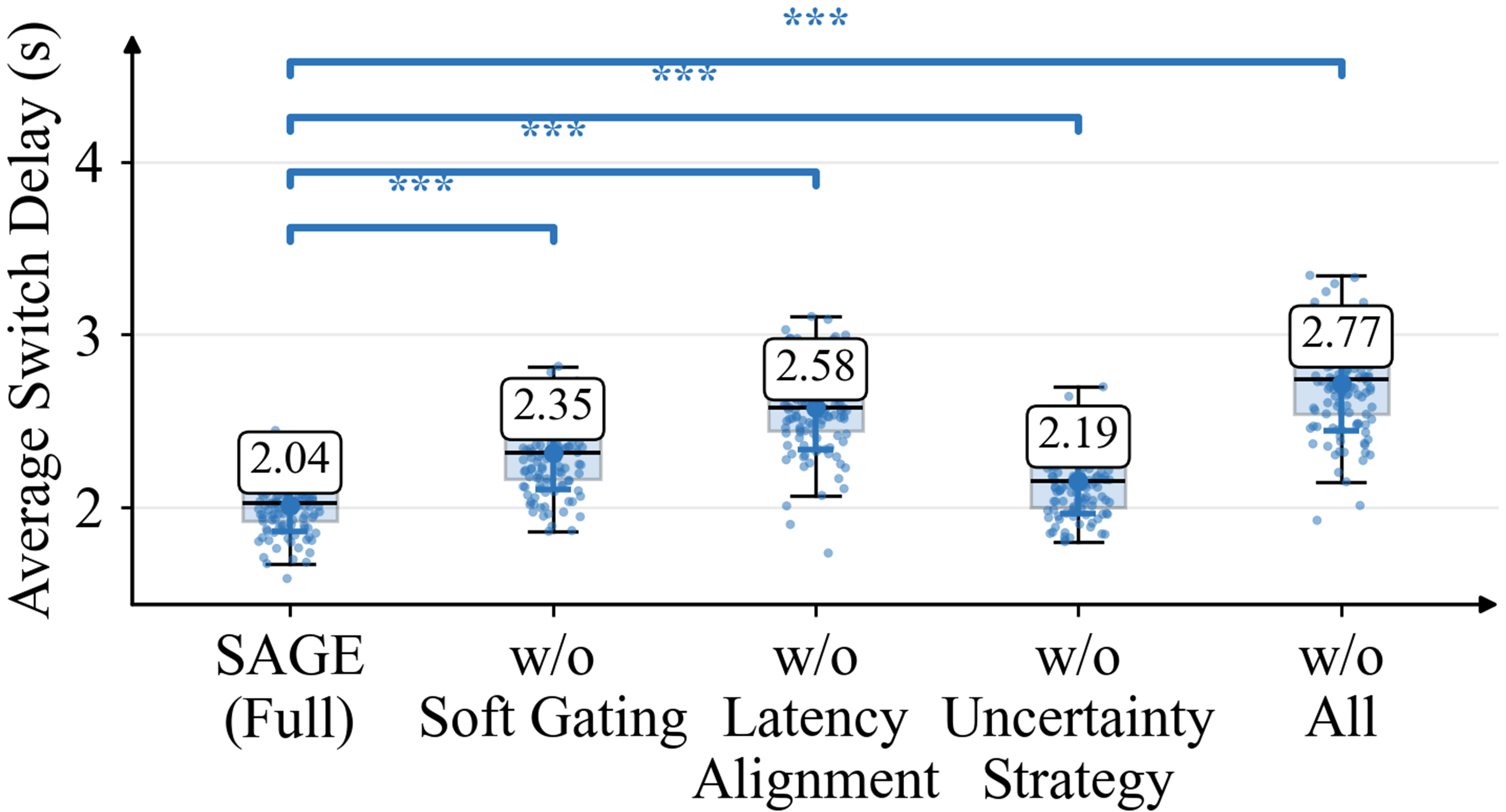}
  \caption{Ablation analysis of average switch latency. ($^{***}$: p-value $<$ 0.001 vs. SAGE Full)}
  \label{fig:fig2}
\end{figure}

To evaluate the contributions of each component in the proposed method, we conducted ablation experiments by removing the switch-aware soft gating, latency-compensated alignment, and uncertainty-driven conservative strategy modules, as well as a variant where all modules were removed.

\begin{table}[!ht]
\renewcommand\arraystretch{1.25}
\caption{Ablation study of the proposed SAGE on the spontaneous attention-switching dataset. }
\centering
\scalebox{0.9}{
\begin{tabular}{lccc}
\hline\hline
\textbf{Model} & \textbf{SI-SDR (dB)} & \textbf{STOI (\%)} & \textbf{ ACC (\%)} \\ \hline
SAGE (Full) & \textbf{8.67} & \textbf{88.24} & \textbf{78.02} \\ 
--w/o Soft Gating & 8.12 & 87.35 & 73.58 \\
--w/o Latency Alignment & 7.86 & 86.02 & 71.34 \\
--w/o Uncertainty Strategy & 7.41 & 85.18 & 74.26 \\
--w/o All & 6.73 & 83.80 & 66.47 \\ \hline\hline
\end{tabular}}
\label{tab:ablation_study_full}
\end{table}

Table~\ref{tab:ablation_study_full} presented the performance of each configuration in terms of SI-SDR, STOI, and switch detection accuracy. Removing any module resulted in performance degradation, highlighting the importance of each design choice. Specifically, removing the soft gating mechanism reduced switch detection accuracy from 78.02\% to 73.58\%, indicating its role in smoothing switching transitions. Removing the latency-compensated alignment module further decreased accuracy to 71.34\%, confirming its importance in reducing the temporal mismatch between EEG responses and switching events for timely tracking. Lastly, removing the uncertainty-driven conservative strategy lowered accuracy to 74.26\%, demonstrating its role in suppressing unreliable decisions during uncertain EEG conditions.
Figure.~\ref{fig:fig2} illustrated the impact of each module on switch delay. Removing the latency alignment and soft gating mechanisms increased switch delay (2.58 s and 2.35 s, respectively), while removing the uncertainty-driven conservative strategy had a smaller effect on delay but significantly reduced speech extraction quality and robustness. Overall, the ablation results verified that each component plays a crucial role in improving speech quality, reducing latency, and enhancing robustness in dynamic target-speaker switching scenarios.

\section{Conclusion}
In this work, we propose SAGE, a switch-aware EEG-guided soft gating framework designed for dynamic target speaker extraction under in-trial attention switching in multi-speaker environments. SAGE combines a dual-stream speech separator with a switch-aware soft gating module, leveraging adaptive temperature scaling to suppress artifacts and ensure smooth transitions during attention switches. It further incorporates latency-compensated alignment with learnable temporal shifts to address label–switch latency gaps, and an uncertainty-driven conservative strategy to stabilize decisions under noisy, unreliable EEG segments. Experimental results on our spontaneous attention-switching dataset show that SAGE consistently outperforms baselines, achieving 8.67 dB SI-SDR and 88.24\% STOI, while reducing the average switch latency to 2.04 s. Ablation studies confirm complementary gains from each component, improving both extraction quality and switching stability. Future work will explore cross-subject generalization and robustness under more diverse acoustic conditions.

\section{Acknowledgments}

This work was supported by the National Key Research and Development Program of China (2025YFF0518003, 2023YFF1203502), the National Natural Science Foundation of China (62371217), and the Center for Computational Science and Engineering at Southern University of Science and Technology.

\section{Generative AI Use Disclosure}
All (co-)authors are responsible and accountable for the work and content of the paper, and consent to its submission. Generative AI tools were used for editing and polishing manuscripts, but were not used for producing a significant part of the manuscript.

\bibliographystyle{IEEEtran}
\bibliography{mybib}

\textbf{}

\end{document}